\documentclass{svproc}
\usepackage{url}
\usepackage{mathrsfs}
\usepackage{hyperref}
\usepackage{amsmath}
\usepackage{amsfonts}
\usepackage{amssymb}
\usepackage{graphicx}
\usepackage{multirow}
\usepackage{hhline}

\begin{document}
\mainmatter              
\title{Bianchi-I cosmology in quantum gravity}
\titlerunning{Bianchi type-I cosmology}  
%
\author{Sunandan Gangopadhyay\inst{1},  Rituparna Mandal \and Amitabha Lahiri}
\authorrunning{Sunandan Gangopadhyay et al.} 
%
\tocauthor{Sunandan Gangopadhyay, Rituparna Mandal and Amitabha Lahiri }
\institute{Department of Theoretical Sciences, S.N. Bose National Centre for Basic Sciences, Block JD, Sector III, Salt Lake, Kolkata 700106, India\\
\email{ sunandan.gangopadhyay@bose.res.in, sunandan.gangopadhyay@gmail.com}}

\maketitle              

\begin{abstract}

The exact renormalization group flow equations for gravity lead to quantum corrections of Newton's constant and cosmological constant. Using this we investigate the Bianchi-I cosmological model at late times. In particular, we obtain the scale factors in different directions, and observe that they eventually evolve into Friedmann-Lema\^itre-Robertson-Walker (FLRW) universe for radiation. However, for stiff matter   the universe shows a Kasner like behaviour. The presentation is based on our work published in \cite{Mandal:2019xlg}.

\keywords{Exact renormalization group equations, Bianchi-I model}
\end{abstract}
%

\section{Introduction}
In the last couple of decades, exact renormalization group (RG) equations \cite{Wetterich:2002yh} have become a powerful tool for nonperturbative investigation of both renormalizable and effective quantum field theories. The framework for exact RG of Euclidean quantum gravity \cite{Reuter:1996cp},\cite{Reuter:2019book} has opened up the possibility of investigating cosmological models \cite{Bonanno:2001xi} in a systematic manner.
An essential ingredient of this framework is the effective average action $\Gamma_{k} [g_{\mu \nu}]$~\cite{Niedermaier:2006wt,Codello:2008vh,Reuter:2001ag,Falls:2014tra}  describing all gravitational phenomena, including the effect of all loops, at a momentum scale $k$\,. Assuming that there is an ultra-violet fixed point much higher than $k$, all quantum fluctuations 
with momenta $p^2 > k^2$ are included, while the  momenta $p^2< k^2$ are suppressed by an infrared regulator. The form of this effective average action is  
\begin{eqnarray}
\Gamma_{k}[g,\bar{g}]=\left(16\pi G(k)\right)^{-1} \int d^{d}x \sqrt{g}\left\lbrace -R(g)+2{\Lambda}(k)\right\rbrace + S_{gf}[g,\bar{g}],
\label{EA}
\end{eqnarray}
where $\bar{g}_{\mu \nu}$ is a background metric and $S_{gf}[g,\bar{g}]$ is a classical background gauge fixing term. The scale dependent Newton's constant $G(k)$ and the cosmological constant $\Lambda(k)$ can be obtained from the RG equation~\cite{Mandal:2019xlg}
\begin{align}
G(k) &= G _{0} \left [ 1-\omega  G_{0} k^{2} + \omega_{1}G^{2}_{0}k^{4}+\mathcal{O}(G^{3}_{0}k^{6}) \right ]\, \label{G} \\ 
\Lambda(k) & =  \Lambda_{0} + G_{0} k^{4} \left[\nu  +\nu _{1}G_{0}k^{2}+\mathcal{O}(G^{2}_{0}k^{4})\right]\,.
\label{flamda}
\end{align}
\section{Bianchi I universe with running $G$ and $\Lambda$}\label{Bianchi}
We now proceed to investigate the Bianchi type-I cosmology which reads  
\begin{eqnarray}
ds^2=-dt^{2}+a^{2}(t) dx^{2}+b^{2}(t) dy^{2}+c^{2}(t) dz^{2}\,.
\label{bianchi1}
\end{eqnarray}
The improved Einstein field equations with time varying Newton's gravitational constant and cosmological constant read 
\begin{eqnarray}
R_{\mu \nu}-\frac{1}{2}Rg_{\mu \nu}=-8\pi G(t)T_{\mu \nu}+\Lambda(t)g_{\mu \nu}
\label{EE}
\end{eqnarray}
with $T_{\mu \nu}=(p+\rho)v_{\mu}v_{\nu}+p g_{\mu \nu}$ being the energy-momentum tensor of a perfect fluid.
The covariant conservation of the energy-momentum tensor 
yields the consistency equation
$8\pi \rho \dot{G}+ \dot{\Lambda}=0.$

One possibility is to relate the scale $k$ to $t$ as $k \sim t^{-1}$ in an FLRW universe~\cite{Bonanno:2001xi}.
For a Bianchi-I universe one needs to include higher order terms in $t^{-1}$. 
We have taken
$k=\frac{\xi}{t}+\frac{\sigma}{t^{2}}+\frac{\delta}{t^{3}}\,.$

\noindent Then we can write from eq.(s) (\ref{G}, \ref{flamda})
%
\begin{align}
G(t) &= G _{0}\left [ 1-\frac{\tilde{\omega}G _{0}}{t^{2}}\left (1+\frac{2\tilde{\sigma }}{t}
+ \frac{2\tilde{\delta }}{t^{2}}+\frac{\tilde{\sigma }^{2}}{t^{2}} \right )+ \frac{\tilde{\omega_{1}} G _{0}^{2}}{t^{4}}+\mathcal{O} \left(\frac{t_{Pl}^{6}}{t^{6}} \right ) \right ]\,, \label{18} \\ \Lambda(t) &= \Lambda _{0}+ \frac{G_{0}}{t^{4}}\left[\tilde{\nu }\left (1+\frac{4\tilde{\sigma }}{t}+ \frac{4\tilde{\delta }}{t^{2}}+\frac{6\tilde{\sigma }^{2}}{t^{2}} \right ) +\frac{\tilde{\nu _{1}}G_{0}}{t^{2}}+\mathcal{O} \left(\frac{t_{Pl}^{4}}{t^{4}} \right ) \right]\,,
\label{19}
\end{align}
where we have defined $\tilde{\omega} \equiv \omega \xi^{2},~\tilde{\omega_{1}} \equiv \omega_{1} \xi^{4},~\tilde{\nu} \equiv \nu \xi^{4}$, $\tilde{\nu_{1}} \equiv \nu_{1} \xi^6$, $\tilde{\sigma} \equiv \frac{\sigma}{\xi}$, $\tilde{\delta} \equiv \frac{\delta}{\xi}$, and $t_{Pl}=\sqrt{G_{0}}$ is the Planck time in natural units. 

For an equation of state $p(t)=\Omega \rho(t)$, the conservation 
of energy-momentum  together with the consistency equation fixes the energy density $\rho$ and the average scale factor ${\mathcal{ R}}=\sqrt[3]{abc}$ to be
\begin{align}
\rho(t)&= \frac{1}{4\pi}\left(\frac{\tilde{\nu}}{\tilde{\omega }}\right)\frac{1}{G_{0}t^{2}}\left \{ 1+\frac{2\tilde{\sigma }}{t}+ \frac{2\tilde{\delta }}{t^{2}}+\frac{\tilde{\sigma }^{2}}{t^{2}}+\left(\frac{2\tilde{\omega_{1}}}{\tilde{\omega}}+\frac{3\tilde{\nu_{1}}}{2\tilde{\nu}} \right)\frac{G_{0}}{t^{2}}+\mathcal{O}\left ( \frac{t_{Pl}^{4}}{t^{4}} \right)\right \}\,, 
\label{26} 
\\
\mathcal{ R}(t)&=\left [ \frac{\mathcal{M} G_{0}}{2}\left ( \frac{\tilde{\omega}}{\tilde{\nu}} \right )\right ]^{\frac{1}{(3+3\Omega)}} t^{\frac{2}{(3+3\Omega)}} \left \{ 1-\frac{1}{(3+3\Omega)}\left (\frac{2\tilde{\sigma }}{t}+ \frac{2\tilde{\delta }}{t^{2}}-\frac{(3\Omega+5)}{(3+3\Omega)}\frac{\tilde{\sigma }^{2}}{t^{2}}+\right. \right. \nonumber \\ & \left. \left. \qquad\qquad\qquad\qquad\qquad\qquad\qquad\qquad\qquad \left(\frac{2\tilde{\omega_{1}}}{\tilde{\omega}}+\frac{3\tilde{\nu_{1}}}{2\tilde{\nu}} \right)\frac{G_{0}}{t^{2}}\right )+\mathcal{O}\left ( \frac{t_{Pl}^{4}}{t^{4}} \right)\right \}
. 
\label{27}
\end{align}
We now discuss two special cases of cosmic matter separately, namely, radiation ($\Omega=\frac{1}{3}$) and stiff fluid ($\Omega=1$). 

\noindent For $\Omega =1/3$, the directional Hubble parameters read,
\begin{eqnarray}
\frac{\dot{a}}{a}&= H(t) + \frac{l(2+\beta)\alpha }{3}H_{1}(t) \,,
\notag\\ 
\frac{\dot{b}}{b} &=  H(t) + \frac{l(\beta-1)\alpha }{3}H_{1}(t) \,,
\notag\\ 
\frac{\dot{c}}{c} &=  H(t) - \frac{l(1+2\beta)\alpha }{3}H_{1}(t)\,. 
\label{34} 
\end{eqnarray}
where $\alpha = \left [ \frac{\mathcal{M} G_{0}}{2}\left ( \frac{\tilde{\omega}}{\tilde{\nu}} \right )\right ]^{-\frac{1}{1+\Omega}}$ and $\beta, l$ are constants of integration.
Here the ``average Hubble parameter'' $H(t)=\dot{\mathcal{R}}/{\mathcal{R}}$ includes isotropic quantum corrections,
\begin{equation}
H(t)=\frac{1}{2t}\left [ 1+\frac{\tilde{\sigma }}{t}+\left ( 2\tilde{\delta}-\tilde{\sigma }^{2}+\left(\frac{2\tilde{\omega_{1}}}{\tilde{\omega}}+\frac{3\tilde{\nu_{1}}}{2\tilde{\nu}} \right)G_{0} \right )\frac{1}{t^{2}} 
+\mathcal{O}\left (\frac{t_{Pl}^{3}}{t^{3}}  \right ) \right ].
\label{Hubble-avg}
\end{equation}
The function $H_1(t)$\, also includes quantum corrections but it will become irrelevant as we shall see now.
Keeping terms up to $\mathcal{O}\left(\frac{t_{Pl}}{t}\right)^4$ 
in the $tt$-component of Einstein equation
and comparing inverse powers of $t$, we find $\Lambda_{0} = 0$,  
$\frac{\tilde{\omega}}{\tilde{\nu}}=8/3$, and $4\left ( \frac{\tilde{\nu }}{\tilde{\omega }}\right )\tilde{\sigma }=3\tilde{\sigma}/2-l^2 \alpha ^{2}\left(\beta^{2}+\beta+1 \right)/3$, leading to 
\begin{equation}
l^2 \alpha ^{2}\left(\beta^{2}+\beta+1 \right)=0\,.
\label{041}
\end{equation}
If $l\neq 0$, we must have $ \beta^{2}+\beta+1 =0 $ which implies that the two roots of $\beta$ are complex. 
Since the scale factors must be real, it follows that $l=0$\,. 
Hence we observe that all the directional Hubble parameters must be equal, that is, the universe must be FLRW in the presence of radiation. 
Thus we conclude that the scale factors of anisotropic Bianchi-I metric flow to the isotropic FLRW cosmology due to RG flow of the Newton's gravitational constant and the cosmological constant. 
%

For $\Omega=1$, which corresponds to stiff matter, the directional Hubble parameters have the same form as in eq. \eqref{34}.
The expression for the (isotropic) average Hubble parameter reads
\begin{eqnarray}
\frac{\dot{\mathcal{R}}}{\mathcal{R}}=\frac{1}{3t}\left [ 1+\frac{\tilde{\sigma }}{t}+\left ( 2\tilde{\delta}-\tilde{\sigma }^{2}+\left(\frac{2\tilde{\omega_{1}}}{\tilde{\omega}}+\frac{3\tilde{\nu_{1}}}{2\tilde{\nu}} \right)G_{0} \right )\frac{1}{t^{2}}+\mathcal{O}\left (\frac{t_{Pl}^{3}}{t^{3}}  \right ) \right ].
\label{51}
\end{eqnarray}
Proceeding as before, we find
%
$\Lambda_{0} = 0\,$
and
\begin{align}
2\left ( \frac{\tilde{\nu}}{\tilde{\omega}} \right ) = \frac{1}{3}\left( 1-\alpha^{2}l^2( \beta^{2}+\beta+1 ) \right).
\label{55}
\end{align}
Writing $l$ in terms of the constant $\beta$, we get
\begin{eqnarray}
l &=& \frac{1}{\alpha}\frac{\sqrt{1-6(\frac{\tilde{\nu}}{\tilde{\omega}})}}{\sqrt{ \left ( \beta ^{2}+\beta +1 \right )}}~\,.
\label{57}
\end{eqnarray}
From the reality of the directional Hubble parameters, we get the following condition from the above equation 
\begin{equation}
1-6\left(\frac{\tilde{\nu}}{\tilde{\omega}}\right) \geq  0  \qquad 
\Rightarrow \qquad \xi^2 \leq  \frac{1}{6} \frac{\omega}{\nu}~.
\label{64}
\end{equation}
When eq. \eqref{64} is an equality, we see from eq. \eqref{55} that $l^2( \beta^{2}+\beta+1 )=0$\,, which implies that $l=0$ since $\beta$ must be real. Therefore, in that case we again have an FLRW universe.

When eq. \eqref{64} is not an equality, that is, $l \neq 0$, we consider the directional Hubble parameters for large $\beta$.
This is  a Kasner type universe, which means that there are expanding and contracting directions. 

\section{Conclusions}

The presentation is based on the study of the Bianchi-I cosmological model at late times taking quantum gravitational effects into account \cite{Mandal:2019xlg}. We used the RG improved cosmological equations following from
the scale dependence of Newton's constant and the cosmological constant.
For radiation, the scale factors take the same form and expand at the same rate in all directions. For stiff matter there are solutions which do not flow to the FLRW universe and show a Kasner like behaviour. The observation of the quantum gravity induced time dependence of the Hubble parameter or scale dependence of $G$, $\Lambda$ would determine the validity of this approach.

%
%

\end{document}